\relax
\documentclass[letterpaper]{article}
\usepackage{aaai17}
\usepackage{times}
\usepackage{helvet}
\usepackage{amsmath}
\usepackage{courier}
\usepackage{graphicx}
\usepackage{array,multirow}
\usepackage{amsmath}
\PassOptionsToPackage{hyphens}{url}
\usepackage{hyperref}
\usepackage{wasysym}
\usepackage{array,colortbl,multirow,multicol,booktabs,ctable}

\begin{document}

\title{Face-to-BMI: Using Computer Vision to Infer Body Mass Index on Social Media\thanks{This is a preprint of a short paper accepted at ICWSM'17. Please cite that version instead.}}
\author{
Enes Kocabey$^{*}$, Mustafa Camurcu$^{\dagger}$, Ferda Ofli$^{\ddagger}$, Yusuf Aytar$^{*}$, Javier Marin$^{*}$, Antonio Torralba$^{*}$, Ingmar Weber$^{\ddagger}$\\$^{*}$ MIT-CSAIL, $^{\dagger}$ Northeastern University, $^{\ddagger}$ Qatar Computing Research Institute, HBKU\\$^{*}$\{kocabey,yaytar,jmarin,torralba\}@mit.edu, $^{\dagger}$camurcu.m@husky.neu.edu, $^{\ddagger}$\{fofli,iweber\}@hbku.edu.qa}

\maketitle

\begin{abstract}
A person's weight status can have profound implications on their life, ranging from mental health, to longevity, to financial income.  At the societal level, ``fat shaming'' and other forms of ``sizeism'' are a growing concern, while increasing obesity rates are linked to ever raising healthcare costs. For these reasons, researchers from a variety of backgrounds are interested in studying obesity from all angles.
To obtain data, traditionally, a person would have to accurately self-report their body-mass index (BMI) or would have to see a doctor to have it measured. In this paper, we show how computer vision can be used to infer a person's BMI from social media images. We hope that our tool, which we release, helps to advance the study of social aspects related to body weight.
\end{abstract}

\section{Introduction}





Together with a person's gender, age and race, their weight status is a publicly visible signal that can have profound influence on many aspects of their life. Most obviously, it can affect their health as having a larger BMI is linked to an increased risk of both cardio-vascular diseases and diabetes, though not necessarily in a straight-forward manner \cite{meigs2006body}. However, other aspects of the burden imposed by obesity come in the form of ``fat shaming'' and other forms of ``sizeism''. For example, obesity is related to a lower  income\footnote{\url{http://www.forbes.com/sites/freekvermeulen/2011/03/22/the-price-of-obesity-how-your-salary-depends-on-your-weight/}} and part of the reason seems to be due to weight-based discrimination \cite{puhletal08ijo}. Even among health professionals ``sizeism'' is so prevalent that it has become a health hazard as, when faced with overweight patients, care providers stop to look for alternative explanations for a medical condition~\cite{doi:10.1080/21604851.2016.1213066}.


For these reasons, researchers from a variety of backgrounds are interested in studying obesity from all angles.
To obtain data, traditionally, a person would have to accurately self-report their body-mass index (BMI) or would have to see a doctor to have it measured. 
In this paper, we propose a new pipeline using state-of-the-art computer vision techniques to infer a person's BMI from social media images, such as their profile picture. We show that the performance in distinguishing for a given pair the more overweight person is similar to human performance. 



\section{Related Work}

Being overweight can lead to a range of negative consequences with the most direct ones concerning health. Given this importance, recent studies in psychology and sociology investigate how humans perceive health from profile pictures. \cite{coetzeeetal09perception} showed that facial adiposity (i.e., perception of weight in the face) was a significant predictor of the perceived health. Furthermore, they showed that perceived facial adiposity was significantly associated with cardiovascular health and reported infections, and hence, an important and valid cue to actual health. In a similar study, \cite{hendersonetal16ptrsl} explored the effect of a variety of facial characteristics on humans' health judgment. They found that facial features such as skin yellowness, mouth curvature and shape were correlated positively whereas facial shape associated with adiposity was correlated negatively with impression of health.

In light of these studies, \cite{webermejova16dh} took a crowdsourcing approach to understand health judgments of humans in a body-weight-inference task from profile pictures. However, since the judgment of whether a picture ``is overweight'' is a rather subjective task, their work suffered from the bias of human annotators to falsely equate ``overweight'' with ``abnormal.'' To eliminate such limitations, \cite{wenguo13ivc} showed that it is indeed feasible to some degree to predict BMI from face images automatically using computational techniques. Their approach relied on detecting a number of fiducial points in each face image and computing hand-crafted geometric facial features to train a regression model for BMI prediction. Their dataset, however, comprised exclusively passport-style frontal face photos with clean background, and hence, the performance of their BMI prediction model is uncertain for noisy social media pictures.

\section{Faces with BMI Data}
To ensure that our system works with noisy, often low quality social media pictures, such as profile pictures, we used the set of annotated images from the VisualBMI project\footnote{\url{http://www.visualbmi.com/}}. These images are, in turn, collected from Reddit posts that link to the imgur.com service. Examples of the underlying Reddit posts can be found in the ``progresspics'' sub-Reddit\footnote{\url{https://www.reddit.com/r/progresspics/}}. The VisualBMI dataset comprises a total of 16,483 images containing a pair of ``before'' and ``after'' images, annotated with gender, height and previous and current body weights. We manually went through all of the image URLs, and cropped the faces. We ignored all the images except the ones with two faces, since we only had previous and current body weights. After the manual cleaning process, we were left with 2103 pairs of faces, with corresponding gender, height and previous and current body weights. Then for each pair, we computed the previous BMI and current BMI. The BMI is defined as (body mass in kg) / (body height in m)${}^2$.

\begin{figure}[ht]
\centering
\includegraphics[width=\columnwidth]{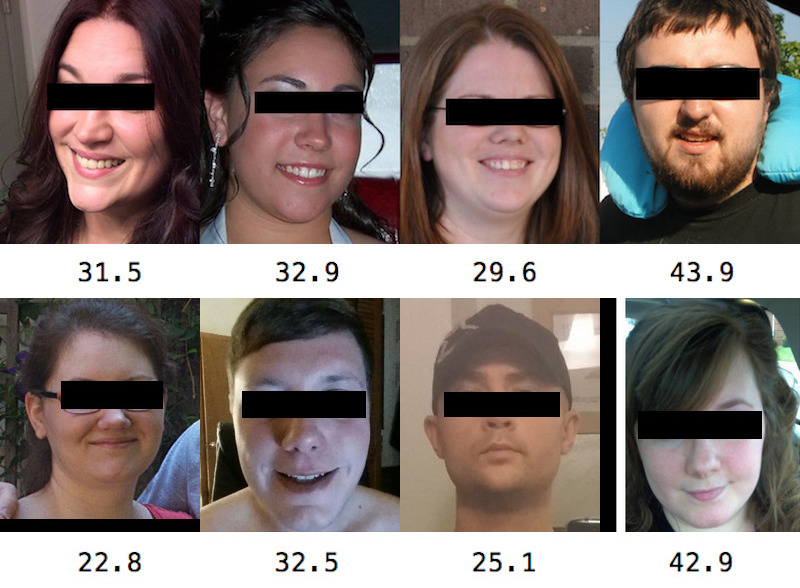}\caption{Examples of the cleaned images used for model training. The black bars have been added to respect user privacy, but the model is learned on the original, public images.}\label{fig:bmiexamples}
\end{figure}

This led to a total of 4206 faces with corresponding gender and BMI information. Of these, seven were in the underweight range (16 $<$ BMI $\le$ 18.5), 680 were normal (18.5 $<$ BMI $\le$ 25), 1151 were overweight (25 $<$ BMI $\le$ 30), 941 were moderately obese (30 $<$ BMI $\le$ 35), 681 were severely obese (35 $<$ BMI $\le$ 40) and 746 were very severely obese (40 $<$ BMI). The dataset contained 2438 males and 1768 females. Figure~\ref{fig:bmiexamples} shows a selection of the faces that were used for training and evaluating our system.

\section{Face-to-BMI System}
In this section, we describe our Face-to-BMI system starting with (i) the computer vision architecture used for building the prediction model, and then (ii) the details of the evaluation and comparison with human performance. Our pre-trained models and scripts for using them can be downloaded for academic research purposes at \url{http://face2bmi.csail.mit.edu}.

\subsection{Computer Vision Architecture}
Many computer vision tasks have greatly benefited from the recent advances in deep learning \cite{Parkhi15,krizhevsky2012imagenet,simonyan2014very} and here we also utilize such models for the Face-to-BMI problem. The features learned in deep convolutional networks are proven to be transferable and quite effective when used in other visual recognition tasks \cite{yosinski2014transferable,girshick2014rich}, particularly when training samples are limited and learning a successful deep model is not feasible due to overfitting. For instance, \cite{ozbulak2016transferable} shows the success of this transfer for age and gender recognition tasks performed on face images. Considering that we also have limited training examples, we adopted a transfer learning approach.  

Our BMI prediction system is composed of two stages: (i) deep feature extraction, and (ii) training a regression model. 
For feature extraction we use two well-known deep models, one trained on general object classification (i.e., VGG-Net \cite{simonyan2014very}) and the other trained on a face recognition task (i.e., VGG-Face \cite{Parkhi15}). Both of these models are deep convolutional models with millions of parameters, and trained on millions of images. The features from the \emph{fc6} layer are extracted for each face image in our training set. For the BMI regression, we use epsilon support vector regression models \cite{smola1997support} due to its robust generalization behavior. The models are trained on the 3368 training images and tested on 838 test images from the VisualBMI dataset. We make sure that the same individual does not exist in both training and test sets (e.g., before and after images). The performance of both of our models are shown in Table~\ref{tab:face2bmi_performance}. As also pointed out by \cite{ozbulak2016transferable}, the features extracted from a more relevant model, i.e., VGG-Face, perform better compared to the VGG-Net features. Due to its superiority we use VGG-Face features in our Face-to-BMI system. 

To see if our system could also be used for tracking weight changes for a single person, rather than comparing across people, we also defined a different train-test split. We first randomly selected 838 unique individuals from our dataset. For each of these individuals, we randomly selected one of the two corresponding before-after face photos and added them to a new test set. All the remaining pictures were added to the new training set. In this way, every person that has a face image in test set also had a face image in the training set. We kept the training and test sizes the same to ensure a fair comparison. Our model achieved 0.68 correlation, compared to 0.65 in the across-people setup, suggesting that our system benefits from having a history of images to train on for the individual it is making a prediction for.


\begin{table}[ht]
\begin{tabular}{lccc}
\toprule
\textbf{Model}   & \textbf{Male} & \textbf{Female} & \textbf{Overall} \\
\midrule
 Face-to-BMI -- VGG-Net  &  0.58 & 0.36 & 0.47 \\
 Face-to-BMI -- VGG-Face &  0.71 & 0.57 & 0.65 \\
\bottomrule
\end{tabular}
\caption{The Pearson $r$ correlations on the test set for the BMI prediction task, broken down by gender. Note that VGG-Face features yield a much better performance.}
\label{tab:face2bmi_performance}
\end{table}
\subsection{Human Evaluation}




We conduct a simple experiment to compare our Face-to-BMI system's performance to that of humans. Given face images for two individuals, each ``contestant,'' i.e., machine and human, is required to tell which one is more overweight. Note that our system was not trained for this specific binary classification task though and a dedicated system might perform better.

For evaluation, we collect a total number of 900 pairs. This is obtained by using only samples coming from the test set, chosen such that pairs are equally distributed in gender subcategories (`male vs.\ male', `female vs.\ female' and `female vs.\ male') and BMI difference between individuals within a pair. Concretely, we randomly collect
300 `male vs.\ male' pairs consisting in 15 subsets of 20-pairs each, such that, for each subset $S_i$, with $i\in\{0,\hdots,14\}$, all pairs $(a,b)\in S_i$ satisfy:
$$(0.5 + i) < |\mathrm{BMI}_a - \mathrm{BMI}_b| \leq (1.5 + i) $$

\noindent We also collect 300 `female vs.\ female' and `female vs.\ male' pairs following the same strategy. In the latter case, we also assure that half of the pairs correspond to males being more overweight and vice-versa. 
Furthermore, we try to balance the overall BMI distribution across the whole spectrum, from thin to obese.

On the human side, we perform the aforementioned questionnaire through Amazon Mechanical Turk\footnote{\url{http://mturk.com}}. Each question is shown to three unique users gathering a total of 2700 answers. The human performance is then obtained using all the answers together and represented as the accuracy. We did not apply a majority voting approach as we wanted to evaluate the performance of an\emph{individual} human. On the machine side, for each question we compare the system output of each individual included in the pair to obtain an answer.  

\begin{figure}[ht]
\centering
\includegraphics[width=\columnwidth]{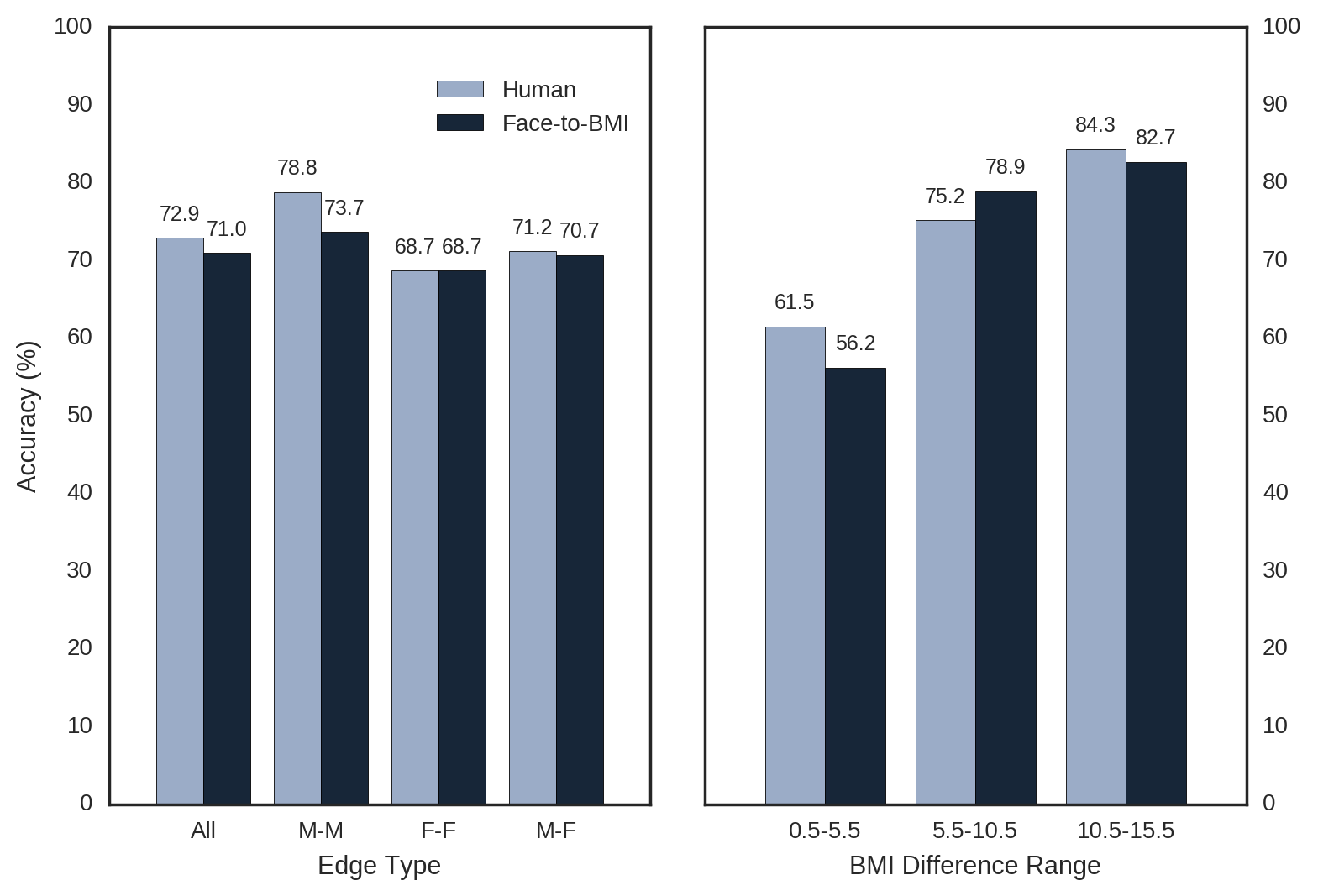}
\caption{Human vs.\ Face-to-BMI comparison broken down by gender and absolute BMI difference between individuals of each pair.}
\label{fig:hvsf2bmi}
\end{figure}

Figure~\ref{fig:hvsf2bmi} depicts the results of comparison broken down by the pair's gender type and the absolute BMI difference between individuals of each pair. The overall performance difference between human and machine is less than $2\%$, but when looking at different gender subcategories, there is a bigger gap the `male vs.\ male' comparisons, $\sim 5\%$. 
Humans slightly outperform the machine for small BMI differences, and there is almost no performance difference for larger BMI differences. 

%




\section{Discussion}



\subsection{Algorithmic Bias}
As demographic groups differ in their BMI distribution, it is likely that, e.g., the race and the BMI are not independent attributes in the visualBMI data. 
This could then mean that our system perpetuates existing stereotypes. For example, as African Americans have higher obesity rates in the US population, an automated system might learn a prior probability that increases the likelihood of a person to be labeled as obese simply based on their race.

To test if our system is biased in outputting a higher BMI for a picture solely due to the person's gender or race, we paired users with a similar BMI, i.e., difference $<$ 1.0, but a different gender or race. Furthermore, these pairs were constructed such that, in aggregate, each demographic group had the same number of (slightly) higher BMIs. For pairs with such close BMIs, an unbiased tool should pick members of either group 50\% of the time. Hence we check if our tool creates a distribution in the output that differs statistically significantly from 50-50. 

For 2000 male-female pairs from the test set the tool predicted a higher BMI for females in 1037 cases, $p=.05$. Though the evidence is inconclusive, it is possible that our system is slightly biased against females. To test for racial bias we did not have a sufficient number of pairs in the test set and hence had to include examples from the training set. For 2000 White-African American pairs, our tool predicted a higher BMI for Whites in 1085 cases, $p<.05$, hinting at a small bias against Whites.




\subsection{Ethical Considerations}
Historically, 19th century phrenology\footnote{\url{https://en.wikipedia.org/wiki/Phrenology}} studied the potential link between the shape of skull and moral attributes, applying a pseudo-scientific methodology to justify racism. Recently, researchers in China have started to predict if a face belongs to a criminal 
which, they claim, they can detect with better-than-random accuracy. However, their work is largely being viewed both as unethical and as scientifically flawed \cite{biddle2016}.

Going beyond moral attributes and the shape of the skull, psychologists have indeed shown that using only facial information it is possible for a person to perform better than random chance at guessing another's personality \cite{BJOP:BJOP254}, an observation at the heart of physiognomy\footnote{\url{https://en.wikipedia.org/wiki/Physiognomy}}. Over the last couple of years, there has been a growing body of work that successfully applies computer vision techniques to automatically infer a person's personality in particular from images shared on social media \cite{Nie2016,Dhall2016,Guntuku:2015:OPY:2813524.2813528,DBLP:conf/icwsm/LiuPSMU16}. 
%
%

Most of the methods mentioned above work ``better than random guessing'' but, when applied to a single individual, are still highly unreliable. This is partly because concepts such as personality are inherently vague and partly because the connection to facial features is weak.

This caveat also largely applies to our tool which, despite a performance similar to humans, is still noisy at the individual level. However, at the \emph{population} level it can be used to detect relative trends as long as it is not biased in a systematic way. In other words, result of the form ``the average BMI for group X is larger than for group Y'' are far more robust than results of the form ``this individual from group X has a higher BMI than this other individual from group Y''. This distinction also applies to the BMI which is useful for studying population health but has shortcomings when used as a tool for individual health \cite{daniels2009use,prentice2001beyond}.

\section{Conclusions}
In this work, we apply the most recent computer vision techniques to obtain a novel Face-to-BMI system. The performance of this tool is on par to that of humans for distinguishing the more overweight person when presented with a pair of profile images. We discuss issues related to algorithmic bias and ethical considerations when inferring information from a person's profile image. To limit the potential of abuse while allowing others to replicate and build on our results, we make our pre-trained models only available to academic researchers after describing the intended use.

In future work, we will apply our method to social media profile pictures to model population-level obesity rates. Preliminary results show that both regional and demographic differences in BMI are reflected in large amounts of Instagram profile pictures.
\bibliography{face_to_bmi}

\begin{thebibliography}{}

\bibitem[\protect\citeauthoryear{Biddle}{2016}]{biddle2016}
Biddle, S.
\newblock 2016.
\newblock Troubling study says artificial intelligence can predict who will be
  criminals based on facial features.
\newblock {\em The Intercept}.
\newblock
  \url{https://theintercept.com/2016/11/18/troubling-study-says-artificial-intelligence-can-predict-who-will-be-criminals-based-on-facial-features/}.

\bibitem[\protect\citeauthoryear{Chrisler and
  Barney}{2016}]{doi:10.1080/21604851.2016.1213066}
Chrisler, J.~C., and Barney, A.
\newblock 2016.
\newblock Sizeism is a health hazard.
\newblock {\em Fat Studies} 0(0):1--16.

\bibitem[\protect\citeauthoryear{Coetzee and
  others}{2009}]{coetzeeetal09perception}
Coetzee, V., et~al.
\newblock 2009.
\newblock Facial adiposity: A cue to health?
\newblock {\em Perception} 38(11):1700--1711.

\bibitem[\protect\citeauthoryear{Daniels}{2009}]{daniels2009use}
Daniels, S.~R.
\newblock 2009.
\newblock The use of bmi in the clinical setting.
\newblock {\em Pediatrics} 124(Supplement 1):S35--S41.

\bibitem[\protect\citeauthoryear{Dhall and Hoey}{2016}]{Dhall2016}
Dhall, A., and Hoey, J.
\newblock 2016.
\newblock First impressions - predicting user personality from twitter profile
  images.
\newblock In {\em HBU},  148--158.

\bibitem[\protect\citeauthoryear{Girshick and others}{2014}]{girshick2014rich}
Girshick, R., et~al.
\newblock 2014.
\newblock Rich feature hierarchies for accurate object detection and semantic
  segmentation.
\newblock In {\em CVPR},  580--587.

\bibitem[\protect\citeauthoryear{Guntuku and
  others}{2015}]{Guntuku:2015:OPY:2813524.2813528}
Guntuku, S.~C., et~al.
\newblock 2015.
\newblock Do others perceive you as you want them to?: Modeling personality
  based on selfies.
\newblock In {\em ASM},  21--26.

\bibitem[\protect\citeauthoryear{Henderson and
  others}{2016}]{hendersonetal16ptrsl}
Henderson, A.~J., et~al.
\newblock 2016.
\newblock Perception of health from facial cues.
\newblock {\em Philosophical Transactions of the Royal Society of London B:
  Biological Sciences} 371(1693).

\bibitem[\protect\citeauthoryear{Krizhevsky and
  others}{2012}]{krizhevsky2012imagenet}
Krizhevsky, A., et~al.
\newblock 2012.
\newblock Imagenet classification with deep convolutional neural networks.
\newblock In {\em NIPS},  1097--1105.

\bibitem[\protect\citeauthoryear{Little and Perrett}{2007}]{BJOP:BJOP254}
Little, A.~C., and Perrett, D.~I.
\newblock 2007.
\newblock Using composite images to assess accuracy in personality attribution
  to faces.
\newblock {\em British Journal of Psychology} 98(1):111--126.

\bibitem[\protect\citeauthoryear{Liu and
  others}{2016}]{DBLP:conf/icwsm/LiuPSMU16}
Liu, L., et~al.
\newblock 2016.
\newblock Analyzing personality through social media profile picture choice.
\newblock In {\em ICWSM},  211--220.

\bibitem[\protect\citeauthoryear{Meigs and others}{2006}]{meigs2006body}
Meigs, J.~B., et~al.
\newblock 2006.
\newblock Body mass index, metabolic syndrome, and risk of type 2 diabetes or
  cardiovascular disease.
\newblock {\em The Journal of Clinical Endocrinology \& Metabolism}
  91(8):2906--2912.

\bibitem[\protect\citeauthoryear{Nie and others}{2016}]{Nie2016}
Nie, J., et~al.
\newblock 2016.
\newblock Social media profiler: Inferring your social media personality from
  visual attributes in portrait.
\newblock In {\em PCM},  640--649.

\bibitem[\protect\citeauthoryear{Ozbulak and
  others}{2016}]{ozbulak2016transferable}
Ozbulak, G., et~al.
\newblock 2016.
\newblock How transferable are cnn-based features for age and gender
  classification?
\newblock In {\em BIOSIG},  1--6.

\bibitem[\protect\citeauthoryear{Parkhi and others}{2015}]{Parkhi15}
Parkhi, O.~M., et~al.
\newblock 2015.
\newblock Deep face recognition.
\newblock In {\em British Machine Vision Conference}.

\bibitem[\protect\citeauthoryear{Prentice and Jebb}{2001}]{prentice2001beyond}
Prentice, A.~M., and Jebb, S.~A.
\newblock 2001.
\newblock Beyond body mass index.
\newblock {\em Obesity reviews} 2(3):141--147.

\bibitem[\protect\citeauthoryear{Puhl and others}{2008}]{puhletal08ijo}
Puhl, R.~M., et~al.
\newblock 2008.
\newblock Perceptions of weight discrimination: prevalence and comparison to
  race and gender discrimination in america.
\newblock {\em Int J Obes} 32:992--1000.

\bibitem[\protect\citeauthoryear{Simonyan and
  Zisserman}{2014}]{simonyan2014very}
Simonyan, K., and Zisserman, A.
\newblock 2014.
\newblock Very deep convolutional networks for large-scale image recognition.
\newblock {\em arXiv preprint arXiv:1409.1556}.

\bibitem[\protect\citeauthoryear{Smola and Vapnik}{1997}]{smola1997support}
Smola, A., and Vapnik, V.
\newblock 1997.
\newblock Support vector regression machines.
\newblock {\em NIPS} 9:155--161.

\bibitem[\protect\citeauthoryear{Weber and Mejova}{2016}]{webermejova16dh}
Weber, I., and Mejova, Y.
\newblock 2016.
\newblock Crowdsourcing health labels: Inferring body weight from profile
  pictures.
\newblock In {\em DH},  105--109.

\bibitem[\protect\citeauthoryear{Wen and Guo}{2013}]{wenguo13ivc}
Wen, L., and Guo, G.
\newblock 2013.
\newblock A computational approach to body mass index prediction from face
  images.
\newblock {\em Image and Vision Computing} 31:392--400.

\bibitem[\protect\citeauthoryear{Yosinski and
  others}{2014}]{yosinski2014transferable}
Yosinski, J., et~al.
\newblock 2014.
\newblock How transferable are features in deep neural networks?
\newblock In {\em NIPS},  3320--3328.

\end{thebibliography}
\bibliographystyle{aaai}

\end{document}